\documentclass[%
onecolumn,
superscriptaddress,
amsmath,amssymb,
aps,
prf,
showkeys,
]{revtex4-2}

\usepackage{graphicx}
\usepackage{amsmath}
\usepackage{amssymb}
\usepackage{hyperref}
\usepackage{bm}
\usepackage{xcolor}

\hypersetup{
    colorlinks = true,
    urlcolor   = blue,
    citecolor  = blue,
}

\newcommand\Wo{\mbox{\textit{Wo}}}
\newcommand{\DPhi}{\Delta \phi}
\newcommand{\micron}{~\mu \mathrm{m}}
\newcommand{\StLayer}{\delta_{St}}

\newcommand\SMfig[1]{\textcolor{green!70!black}{ {S}#1}}
\newcommand\SMref[1]{\textcolor{green!70!black}{#1}}

\begin{document}

\title{Energetics, shearing and pumping efficiency of propagating contractions over villi-patterned wall}

\author{Rohan Vernekar}
\email{rohan.vernekar@orange.fr}
\affiliation{Univ. Grenoble Alpes, CNRS, Grenoble INP, Laboratoire Rh\'{e}ologie et Proc\'{e}d\'{e}s (LRP), 38000 Grenoble, France}
\author{Claude Loverdo}
\affiliation{Sorbonne Université, CNRS, Institut de Biologie Paris-Seine (IBPS), Laboratoire Jean Perrin (LJP), F-75005 Paris, France}
\author{Stéphane Tanguy}
\affiliation{Univ. Grenoble Alpes, CNRS, UMR 5525, VetAgro Sup, Grenoble INP, TIMC, 38000 Grenoble, France }
\author{Cl\'{e}ment de Loubens}
\email{clement.de-loubens@univ-grenoble-alpes.fr}
\affiliation{Univ. Grenoble Alpes, CNRS, Grenoble INP, Laboratoire Rh\'{e}ologie et Proc\'{e}d\'{e}s (LRP), 38000 Grenoble, France}

\date{\today}

\begin{abstract}
	Intestinal villi undergo pendular-wave motility---an active, propagating tissue motion driven by underlying longitudinal muscles.
	This motility drives irreversible, counter-wave fluid pumping, akin to the antiplectic metachrony of ciliary carpets, and generates a viscous mixing boundary layer above the villi tips, whose height is controlled by flow inertia.
	Using a simplified 2D model of the rat duodenum, we quantify the system's viscous energy dissipation and axial pumping efficiency.
	In contrast to the classical Stokes’ second problem, we show that the fluid volume dominating energy dissipation is dictated by the intervillous geometry, remaining insensitive to the dynamically varying viscous mixing boundary layer height.
	The computed pumping efficiency is orders of magnitude lower than that of canonical peristalsis for equivalent flux pumping.
	We thus infer that bulk fluid pumping is not the primary biophysical function of propagating pendular-wave motility; instead, we postulate that its main role is to shear the mucus barrier layer over the villi-lined mucosa.
	Comparing the strain rate in the barrier region with canonical peristaltic reference values for a villi-free wall strongly supports our hypothesis.
	Finally, for biomimetic microfluidic applications, geometric optimization reveals that pumping efficiency scales quadratically with the channel-to-villi height ratio in Stokes flow, whereas in the inertial regime, dynamic flux confinement renders this geometric optimization strategy redundant.
\end{abstract}


\maketitle

\section{Introduction}
\label{sec:intro}

Biological organisms interact with their environment through specialized interfaces that mediate mass transport between the living system and its surroundings. 
This exchange is made possible by the presence of active microscopic finger-like structures, such as ciliated cells in corals and lungs, or villi in the small intestine, which induce complex flow patterns~\citep{pacherres_AcuteTemperatureEffects_2026,komatsu_MicroscopicAnalysisAngled_2026,hazra_AerosolDepositionMucuslined_2025,zhao_HarnessingCiliumOscillations_2026,briole2026disentangling}. 
Concurrently, these fundamental physiological insights have stimulated the development of advanced biomimetic microfluidic systems \citep{loganathan_MagneticCiliaProgrammable_2025,skarsetz_SoftRoboticEngines_2025, wang_ElectronicallyActuatedArtificial_2024, sun_OptimisationNetFlow_2025}.

From both physiological and biomimetic perspectives, a central question concerns the energetic cost of these active transport mechanisms relative to their hydrodynamic function, namely whether they primarily promote transport, mixing or local shearing.
Furthermore, how does this cost compare with classically studied pumping strategies such as peristalsis, ubiquitous in biological systems~\citep{jaffrin_PeristalticPumping_1971}?

While the motion of these active bio-interfaces can be highly complex~\citep{lentle_MucosalMicrofoldsAugment_2013, lim_FlowMixingSmall_2015,ishikawa_FluidDynamicsSquirmers_2024, komatsu_MicroscopicAnalysisAngled_2026}, we recently introduced a minimal framework inspired by rat intestinal physiology~\citep{lentle_ComparisonOrganizationLongitudinal_2012} where an array of rigid villi undergoes oscillatory longitudinal motion~\citep{vernekar_HydrodynamicsVillipatternedChannel_2025}. 
By imposing a phase lag between neighbouring villi ($\DPhi$), the emergent \textit{pendular-wave activity} generates both a viscous mixing layer and counter-wave pumping, analogous to antiplectic metachrony in ciliary systems.
Remarkably, these features emerge despite the absence of structural flexibility, a property often used to break the time reversibility of Stokes flows~\citep{brennen_FluidMechanicsPropulsion_1977}.
Instead, the flow is governed by the non-reciprocal collective motion of the villi, in close analogy with the minimalist three-sphere swimmer model of~\citet{najafi_SimpleSwimmerLow_2004}.
With oscillatory inertial effects, both pumping and mixing are non-monotonically modulated by the dynamic confinement of the Stokes length scale, $\StLayer$~\citep{schlichting_BoundaryLayerTheory_1960}.

We leverage this minimal framework to evaluate the fundamental trade-offs between energetic cost and hydrodynamic utility in active micro-patterned systems.
Focusing on the rat-gut derived 2D villi-patterned channel, we construct maps across a broad range of phase lags ($\DPhi$) and oscillation frequencies ($\omega$) to quantify the energetic cost, pumping efficiency, near-villi strain rates, and enstrophy generated by pendular-wave activity.
We find that while the bulk pumping efficiency is several orders of magnitude lower than for peristalsis, the activity generates elevated near-villi strain rates and enstrophy, leading us to infer that localized shearing and homogenisation are its primary physiological roles.
Finally, we demonstrate a simple geometric optimization strategy in the Stokes-flow regime.

\begin{figure}[t]
	\centering
	\includegraphics[width=\linewidth]{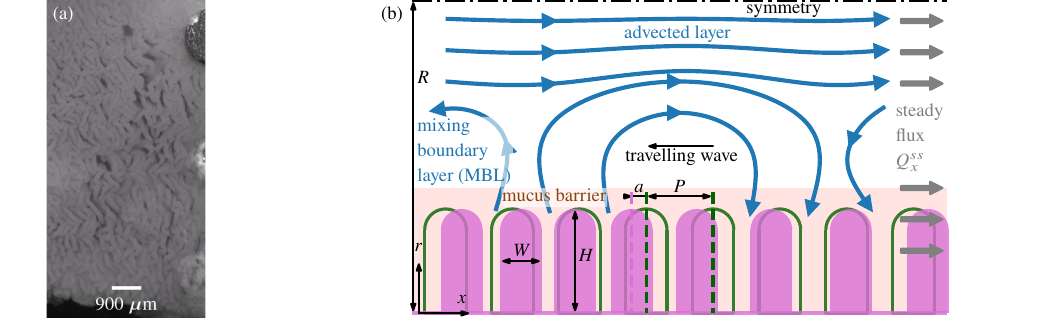}
	\caption{\textbf{(a)} Top view of the ridge or leaf-like villi in the duodenum of the rat. \textbf{(b)} The 2D simulation model with eight axially periodic villi. The mixing boundary layer (MBL) dominated by the instantaneous semi-vortical flow patterns above the villi, and the uniform flow advecting layer above it are illustrated. The villi geometry is defined by the villi width $W$, villi height $H$, intervillous pitch $P$ and oscillation amplitude $a$.}
	\label{fig:illustrations}
\end{figure}

\section{Physical model and numerical method}
\label{sec:method}

\subsection{Two-dimensional model and governing equations}
\label{subsec:model}

As seen in figure~\ref{fig:illustrations}(a), the duodenal villi exhibit a complex, folded ridge-like morphology.
To extract fundamental insights into the fluid mechanics of their motility, we abstract this topography and idealize the rat duodenum as a 2D planar axially periodic channel of depth $2R$ lined with transverse, rigid villous ridges of height $H$ and width $W$, and pitch $P$ (figure~\ref{fig:illustrations}(b)).
The top and bottom walls undergo symmetric, in-phase motility.
Each villus (indexed $n$) oscillates axially with a harmonic velocity,
\begin{equation}
	\label{eq:villiOscVel}
	U^n_i(t) = U_0 \sin \left( \omega t + (n-1) \Delta \phi \right) \hat{x}_i \, ; \quad n=1,2, \dots ,N,
\end{equation}
\noindent
where $U_0=\omega a$ is the maximum oscillation velocity, $\omega=2\pi f$ is the angular frequency, $a$ is the oscillation amplitude, and $\DPhi$ is the constant phase lag between adjacent villi.
A linear velocity profile is imposed along the intervillous gap bottom wall, interpolating between the velocities of the bounding villi.
The channel periodic length is $L_x=NP$, with periodic inflow-outflow applied at the channel axial ends ($\pm x$).
Adopting geometric values approximately corresponding to the rat duodenum, we get $R=5.6H$, $P/W=1.6$, and $H/W = 2.5$ \citep{vernekar_HydrodynamicsVillipatternedChannel_2025}.

The fluid is modelled as Newtonian with dynamic viscosity $\mu$ and density $\rho$.
We non-dimensionalize the Navier-Stokes (NS) equations using characteristic scales for velocity $\tilde{\bm{u}} = \bm{u}/(\omega a)$, pressure $\tilde{p} = p W/(\mu \omega a)$, length $\tilde{\bm{x}} = \bm{x}/W$, and time $\tilde{t}=t \omega$, yielding the dimensionless NS equations,
\begin{equation}
	\label{eq:NSeqs_ND}
	\tilde{\partial}_i \tilde{u}_i = 0 \quad \text{ and } \quad
	\frac{1}{\tilde{a}} \frac{\partial \tilde{u}_i}{\partial \tilde{t}} + \tilde{u}_j  \tilde{\partial}_j \tilde{u}_i = - \frac{1}{\tilde{a} \Wo^2} \tilde{\partial}_i \tilde{p} + \frac{1}{\tilde{a} \Wo^2} \tilde{\partial}_j \tilde{\partial}_j  \tilde{u}_i .
\end{equation}
\noindent
Thus, the system is governed by three non-dimensional parameters, namely, the Womersley number $\Wo = W\sqrt{\omega \rho / \mu}$, the reduced amplitude $\tilde{a} = a/W$, and the phase lag $\DPhi$.

\subsection{Numerical method and convergence}
\label{subsec:lbm}

Simulations are carried out varying $\Wo$ and $\DPhi$, in the physiologically motivated small amplitude limit ($\tilde{a} \le 0.2$).
The fluid flow is simulated using the lattice Boltzmann method (LBM) \citep{vernekar_HydrodynamicsVillipatternedChannel_2025}.
We employ the two-relaxation-time (TRT-D2Q9) scheme \citep{ginzburg_TworelaxationtimeLatticeBoltzmann_2008} alongside the incompressible equilibrium \citep{he_LatticeBoltzmannModel_1997} to accurately capture the second-order time-averaged flow phenomena.
The moving rigid boundaries are resolved using a variation of the two-node interpolated bounce-back (IBB-CLI) scheme \citep{ginzburg_UnifiedDirectionalParabolicaccurate_2023}.
The uncovered fresh nodes are repopulated using the local iteration refill (LIR) procedure \citep{tao_InvestigationMomentumExchange_2016}.
To resolve the geometry, we set the villus and channel dimensions to $W=20$ and $R=280$ in lattice units, respectively.
We simulate varying Womersley numbers by scaling both $\mu$ and $\omega$, while the geometric length ratios are held constant.

Simulations are initialized from rest and run until the flow reaches a time-periodic steady state.
Convergence is determined by monitoring the developing time-averaged steady streaming flow (SSF) velocity (${u}_i^{{ss},\bullet}({x_j})$), evaluated at the end of every oscillation period $T$.
We consider the simulation converged when the ${L_2}$ norm of the relative change in SSF field falls below $1\%$ (see supplementary section~\SMref{1B}).
Once converged, the simulation is run for one additional period, when we record snapshots of the instantaneous velocity field ${u_i}({x_j},t)$, the final converged SSF ${u_i}^{{ss}} ({x_j})$ and the pumped steady flux $Q^{ss}_x$.
The specifics of the numerical implementation and convergence are detailed and validated in \citep{vernekar_HydrodynamicsVillipatternedChannel_2025} for resolving the net axial and radial pumping generated by the pendular-wave motion.

\begin{figure}[t]
	\centering
	\includegraphics[width=\linewidth]{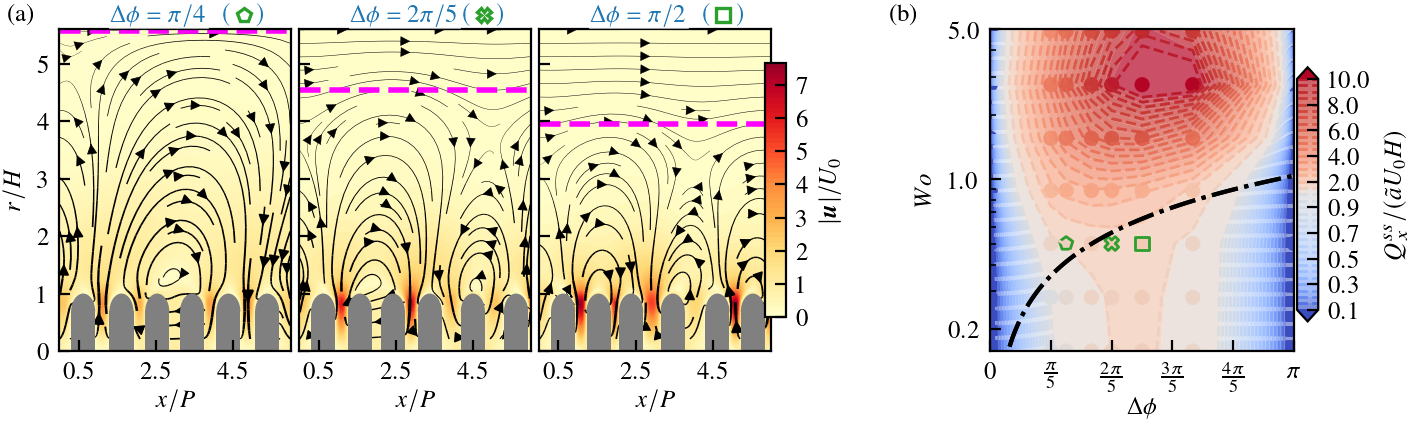}
	\caption{\textbf{(a)} Three panels of instantaneous flow fields at $t/T=0.45$ for increasing phase lags $\DPhi=\pi/4$, $2\pi/5$ and $\pi/2$, respectively. Flow is plotted for $\Wo=0.5$ and $\tilde{a}=0.2$, with streamline width proportional to the velocity magnitude. Dashed (magenta) line shows the approximate separation between the mixing boundary layer (MBL) and the advected layer. \textbf{(b)} Rescaled steady axial pumping flux contours over the coordinates ($\DPhi$, $\Wo$). Dashed line contours for $\tilde{a}=0.1$ are superimposed on filled colour contours for $\tilde{a}=0.2$. Dash-dotted (black) transition curve gives the separation between viscous-dominated (below) and inertia-dominated (above) regimes. Empty (green) markers in \textbf{(b)} indicate parametric location of streamplots in \textbf{(a)}.}
	\label{fig:instantFlow_SSFlux}
\end{figure}

\section{Instantaneous flow-field and steady pumping}
\label{sec:instantFlowfield}
The pendular-wave motility generates a distinct mixing boundary layer (MBL) above the villi tips, characterized by counter-rotating asymmetric semi-vortical structures, and an overlying uniform advected layer (figure~\ref{fig:instantFlow_SSFlux}(a)), reminiscent of cilia driven flow patterns \citep{pacherres_AcuteTemperatureEffects_2026}.
As extensively investigated by \citet{vernekar_HydrodynamicsVillipatternedChannel_2025}, the MBL height $\ell$  grows with decreasing $\DPhi$ and $\Wo$, until it reaches the central symmetry boundary.
Once the effect of the oscillation amplitude $\tilde{a}$ is scaled out, the crossover from viscous-dominated to an inertia-dominated flow occurs at $\Wo \approx 1.2W\DPhi^{2/3}/H$.
While physiological pendular motility in the gut operates firmly within the viscous regime \citep{deloubens_FluidMechanicalConsequences_2013}, the inertial regime is highly relevant to biomimetic microfluidic systems.

This boundary actuation generates an irreversible steady-streaming axial flux directed opposite to the travelling wave.
The time-averaged axial pumping flux in the half-channel is $Q^{ss}_x = \frac{1}{T L_x} \int_{0}^{L_x} \int_{t}^{t+T} \int_{0}^{R} u_x(x,r,t)\, dr \, dt \, dx$.
Figure~\ref{fig:instantFlow_SSFlux}(b) maps this flux over the $(\DPhi, \Wo)$ parameter space.
The flux contours for $\tilde{a}=0.1$ (dashed lines) and $\tilde{a}=0.2$ (filled colours) show excellent collapse after  rescaling by the factor $\tilde{a}U_0 H$.
In the viscous regime, the flux is independent of $\Wo$ and observes the geometric scaling $Q^{ss}_x \approx a U_0 R \sin(\DPhi) / (2P)$ \citep{vernekar_HydrodynamicsVillipatternedChannel_2025}.
However, in the inertial regime, the flux is non-monotonic in both $\Wo$ and $\DPhi$.
Although increasing the oscillatory inertia increases the steady flow velocity scale, this faster flow becomes increasingly confined to a smaller area above the villi tips due to the dynamic confinement imposed by the decreasing MBL ($\ell \propto \Wo^{-1/2}$).
Consequently, the peak flux is reached at moderate inertia ($\Wo \approx 2.8$) before decreasing at higher $\Wo$.

\section{Viscous dissipation}
\label{sec:viscDissip}
\begin{figure}
	\centering
	\includegraphics[width=\textwidth]{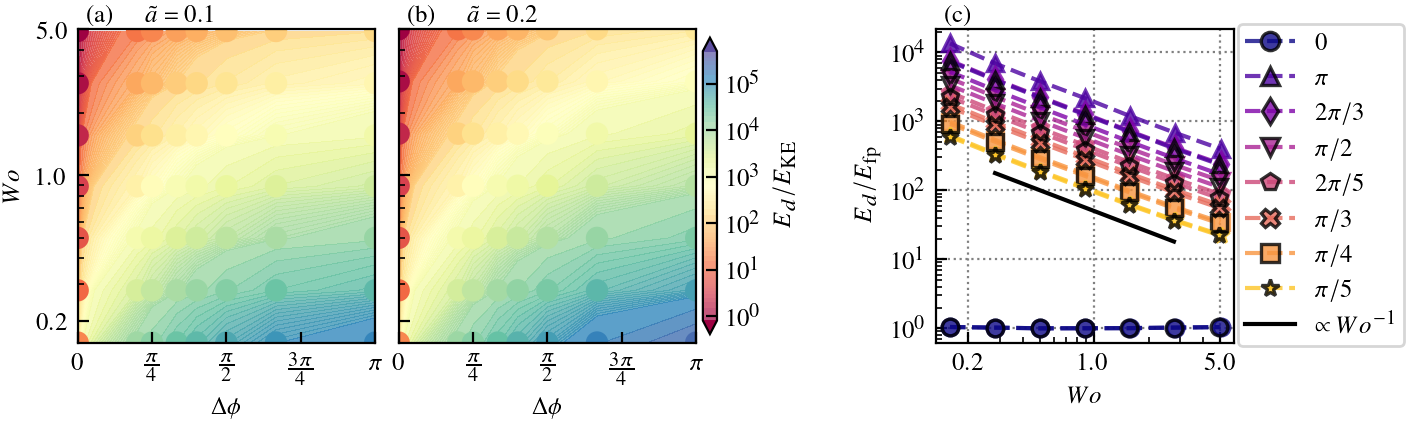}
	\caption{Total viscous dissipation per cycle, $E_d$ scaled by the characteristic kinetic energy imposed by an oscillating villus $E_{KE}$ (which depends on $U_0\propto \omega \propto \Wo^2$), for \textbf{(a)} $\tilde{a}=0.1$ and \textbf{(b)} $\tilde{a}=0.2$. \textbf{(c)} Ratio of $E_d$ to the corresponding total dissipation in the classical oscillating flat-plate problem, $E_{\mathrm{fp}}$. Empty and filled markers correspond to $\tilde{a}=0.1$ and $0.2$, respectively. See text for expressions.}
	\label{fig:viscDissipMaps}
\end{figure}

Since the flow is periodic and driven exclusively by boundary actuation, the total mechanical work performed to pump the fluid is exactly balanced by the viscous dissipation.
The total viscous energy dissipated per cycle over an intervillous pitch $P$ is $E_d = \frac{1}{N} \int_{t}^{t+T} \int_{\Omega} 2\mu S_{ij} S_{ij} \, d\Omega \, dt$, where $S_{ij}$ is the symmetric strain rate tensor and $\Omega$ the fluid domain.
Figures~\ref{fig:viscDissipMaps}(a) and (b) plot the dimensionless dissipation per cycle $\tilde{E}_d$, normalized by the characteristic kinetic energy, $E_{\mathrm{KE}} = \rho V_v U^2_0 / 2$ per single villus of volume $V_v$.
Maximum dissipation occurs at $\DPhi = \pi$ when neighbouring villi move exactly out-of-phase, and decreases monotonically with the phase lag.
For synchronous oscillations ($\DPhi = 0$), the dissipation scales as $\tilde{E}_d \propto \Wo^{-1}$. 
For all asynchronous cases ($\DPhi > 0$), the dissipation is at least an order of magnitude higher, scaling as $\tilde{E}_d \propto \Wo^{-2}$ (see supplementary figure~\SMfig{1}).

In contrast to the axial flux (figure~\ref{fig:instantFlow_SSFlux}(b)), viscous energy dissipation maps remain unaltered by the onset of oscillatory fluid inertia, challenging the naive assumption that dissipation would be determined by the MBL dynamics.
To understand this, we compare $E_d$ with the energy dissipated by a classical oscillating flat plate of length $P$ in figure~\ref{fig:viscDissipMaps}(c), where $E_{\mathrm{fp}} = \left(\pi P \rho U_0^2/2 \right) \sqrt{2 \mu/(\rho \omega)} \propto \Wo^{3}$ \citep{schlichting_BoundaryLayerTheory_1960}.
The scaling shows that $E_{d} \sim E_{\mathrm{fp}}$ for synchronous oscillations ($\DPhi=0$), and $E_{d}  \propto E_{\mathrm{fp}}/\Wo$ for cases with $\DPhi > 0$.
Thus, without intervillous contractions, the dissipative behaviour follows analytical flat-plate theory over the investigated $\Wo$.

The physics of total viscous dissipation can be expressed as the product of fluid volume, dissipated power density, and time period ($T \propto \Wo^{-2}$). 
For the flat-plate theory, the length scale that underpins both the velocity gradients and the dominant volume for dissipation is given by the Stokes layer $\StLayer = \sqrt{\nu/\omega}\propto \Wo^{-1}$.
For $\DPhi > 0$, the MBL exhibits a slower decay of $\ell \propto \StLayer^{1/2} \propto  \Wo^{-1/2}$ \citep{vernekar_HydrodynamicsVillipatternedChannel_2025}. 
If this dynamically varying length scale were to dictate the volume for dissipation, the observed $E_d \propto \Wo^{2}$ scaling would be mathematically impossible.
The robust $\Wo^{2}$ scaling therefore indicates that the relevant length scale for dissipation is set entirely by the villi-wall geometry, independent of the MBL height.
Consequently, for pendular-wave motility, the viscous dissipation within the mixing layer above the villi tips remains subdominant to the energy dissipated within the intervillous spaces.
This deduction also explains why axial flow inertia, which drastically affects the MBL height, leaves no imprint on the total dissipation maps (figures~\ref{fig:viscDissipMaps}(a) and (b)).

\begin{figure}
	\centering
	\includegraphics[width=\linewidth]{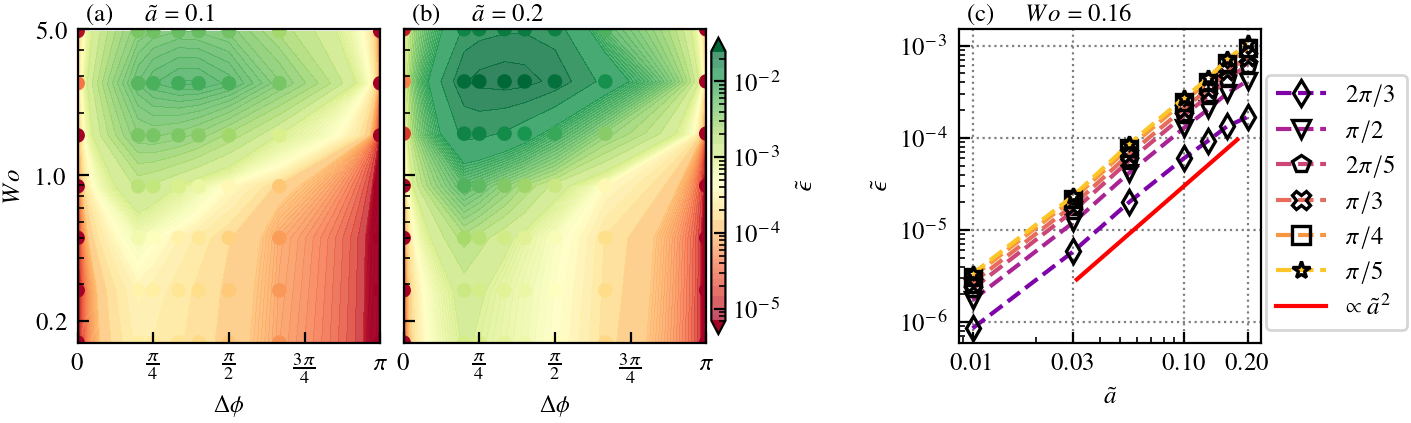}
	\caption{Axial pumping hydrodynamic efficiency maps for simulations with \textbf{(a)} $\tilde{a}=0.1$ and \textbf{(b)} $\tilde{a}=0.2$ (also see supplementary figure~\SMfig{3}). \textbf{(c)} Variation of hydrodynamic efficiency with villi oscillation amplitude in the small-amplitude limit under Stokes flow conditions ($\Wo\approx0.16$).}
	\label{fig:pumpingEfficiencyRatio}
\end{figure}

\section{Axial pumping efficiency}
\label{sec:efficiency}

\subsection{Efficiency parameter}
\label{subsec:efficiencyParam}
To evaluate the energetic cost of fluid propulsion, we introduce a non-dimensional hydrodynamic efficiency parameter, adapted from ciliary metachronal pumping \citep{osterman_FindingCiliaryBeating_2011}, $\tilde{\epsilon} = {\mu \left( Q^{ss}_x \right)^2} / {\left(P H \Phi_d \right)}$, where $\Phi_d = E_d/T$ is the time-averaged dissipated power per villus.
Figures~\ref{fig:pumpingEfficiencyRatio}(a) and (b) map $\tilde{\epsilon}$ over $(\DPhi, \Wo)$.
Unlike total viscous dissipation, increasing oscillatory fluid inertia exerts a distinct non-monotonic effect on  efficiency.
The system reaches a peak efficiency of $\tilde{\epsilon} \approx 2.77\%$ at moderate inertia  ($\Wo=2.8$) and intermediate phase lag ($\DPhi=\pi/3$) for $\tilde{a}=0.2$.
Focusing on the Stokes flow regime ($\Wo = 0.16$) in the limit of small oscillations ($\tilde{a} \le 0.2$), we see an approximate quadratic dependence of the efficiency on $\tilde{a}$ in figure~\ref{fig:pumpingEfficiencyRatio}(c), consistent with viscous regime scaling arguments \citep{vernekar_HydrodynamicsVillipatternedChannel_2025}.

The gut also experiences flow pumping due to peristalsis, which drives fluid by generating propagating pressure gradients.
For comparing pendular-villi pumping with other biophysical pressure-driven mechanisms, an alternative channel pumping efficiency is defined as $\tilde{\eta} = \Phi_{\min} / \Phi_d$.
Here, $\Phi_{\min}$ is the theoretical minimum power dissipated to drive the same flux $Q^{ss}_x$ via ideal Poiseuille flow in the same channel over length $P$, given as $\Phi_{\min} = {3 P \mu} \left(Q^{ss}_x \right)^2 / R^3$ \citep{serrin_MathematicalPrinciplesClassical_1959}.
Contours of $\tilde{\eta}$ follow similar patterns as those for $\tilde{\epsilon}$, with a peak value of $\tilde{\eta} \approx 1.9\times 10^{-4}$ (see supplementary figure~\SMfig{2}).
The biological flow regime in the small intestine operates at low Womersley numbers ($\Wo \lesssim 0.5$), with phase lags considerably smaller than those sampled in our simulations, typically $\DPhi \approx \mathcal{O}(0.1)$ \citep{lentle_ComparisonOrganizationLongitudinal_2012}.
Thus, the physiologically relevant flow is viscous-dominated.
Under these conditions the channel pumping efficiency is remarkably low ($\tilde{\eta} \lesssim 10^{-5}$).

\subsection{Comparison with peristalsis}
\label{sec:efficiency:subsec:peristalsis}
\begin{figure}
	\centering
	\includegraphics[width=\linewidth]{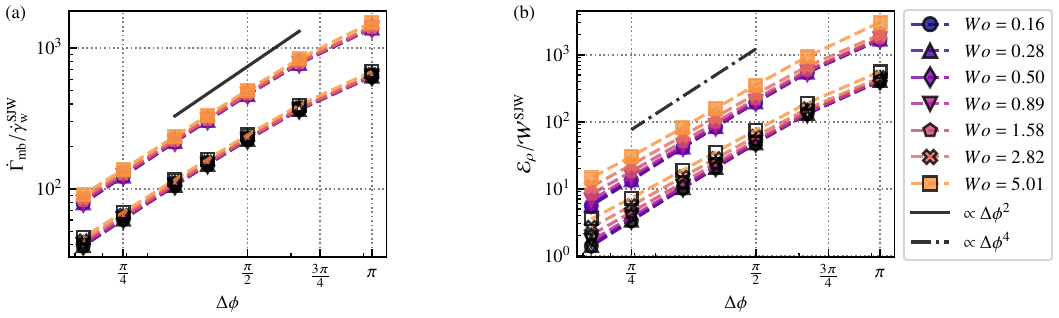}
	\caption{\textbf{(a)} Rescaled strain rate magnitude in the mucus barrier ($0 \le r \le 6H/5$) and \textbf{(b)} rescaled enstrophy density in the lumen above the villi ($r > H$). Open and filled markers denote values for reduced amplitudes $\tilde{a} = 0.1$ and $0.2$, respectively. Reference scales $\dot{\gamma}^{\mathrm{SJW}}_{\mathrm{w}}$ and $\mathcal{W}^{\mathrm{SJW}}$ are derived from the analytical \citeauthor*{shapiro_PeristalticPumpingLong_1969} (SJW) solution for peristalsis.}
	\label{fig:shearAndEnstrophyRatio}
\end{figure}

We contextualize the pumping efficiency values against the better-known peristaltic mechanism using the canonical lubrication-theory solution of Shapiro, Jaffrin and Weinberg (SJW) \citep{shapiro_PeristalticPumpingLong_1969} with the ``free-pumping'' condition.
The peristaltic reference half-channel flux is denoted as $Q^{\mathrm{SJW}}_{x}$, the  leading order time and domain-averaged enstrophy or vorticity reference value as $\mathcal{W}^{\mathrm{SJW}}$, and similarly the time and wavelength-averaged wall strain-rate value as $\dot{\gamma}_{\mathrm{w}}^{\mathrm{SJW}}$ (see supplementary section~\SMref{1D}).
Utilizing \textit{in vivo} parameters for the rat jejunum peristalsis (wave-speed of $c_p\approx4.3$~mm/s, a mean radius $R\approx1.72$~mm, and a relative deflection amplitude $\beta\approx0.58$) \citep{ailiani_QuantitativeAnalysisPeristaltic_2009}, we obtain a theoretical flux of $Q^{\mathrm{SJW}}_x=3.2$~mm$^2/$s.
Thus, the channel efficiency is approximated as $\tilde{\eta} = 3 \left(Q^{\mathrm{SJW}}_x \right)^2 / \left( R^4 \mathcal{W}^{\mathrm{SJW}} \right) \approx 0.98$.

Channel peristalsis thus outperforms pendular-wave pumping by several orders of magnitude.
We therefore infer that bulk fluid pumping is not the primary physiological function of propagating longitudinal motility observed over the intestinal submucosa.
Instead, we postulate that a key biophysical function of this pendular motility is to introduce local shear into the secreted mucus barrier region covering the villi, as well as to induce chaotic mixing in the lumen.
To substantiate this, we define a mucus barrier region $\Omega_{\mathrm{mb}}$ extending $H/5$ (${\approx}100\micron$) above the villi tips (figure~\ref{fig:illustrations}(b)) \citep{paone_MucusBarrierMucins_2020}.
We evaluate the spatiotemporal average of the strain-rate invariant within the mucus barrier region in our fluid simulations as $\dot{\Gamma}_{\mathrm{mb}} = \frac{1}{|\Omega_{\mathrm{mb}}| T} \int_{\Omega_{\mathrm{mb}}}   \int_{t}^{t+T} \sqrt{2 S_{ij} S_{ij}} \, dt \, d\Omega$. Similarly, we compute the averaged enstrophy density in the luminal region extending above the villi tips as $\mathcal{E}_{\rho} = \frac{1}{(R-H) L_x T} \int_{0}^{L_x} \int_{H}^{R}  \int_{t}^{t+T} \left( {\partial_i u_j} - {\partial_j u_i} \right)^2 \, dt \, dr \, dx$.

Figures~\ref{fig:shearAndEnstrophyRatio}(a) and (b) show the mucus barrier strain rate and the luminal enstrophy density rescaled by peristaltic references with identically matched kinematic parameters ($c_p=\omega P / \DPhi$ and $\beta=H/R$), respectively.
The barrier strain rate exceeds the peristaltic reference scale by at least an order of magnitude, scaling quadratically with the phase lag (${\propto} \tilde{a} \DPhi^2$).
The luminal enstrophy density ratio shows elevated values over the sampled range and a scaling dependence ${\propto} \tilde{a}^2 \DPhi^4$.
These elevated kinematic values strongly indicate that while pendular waves are highly inefficient at driving bulk axial flow, our results suggest that their true physiological utility lies in shearing and homogenizing the secreted mucus layer over the villi and enhancing nutrient transport towards the villi for absorption.

\begin{figure}
	\centering
	\includegraphics[width=\linewidth]{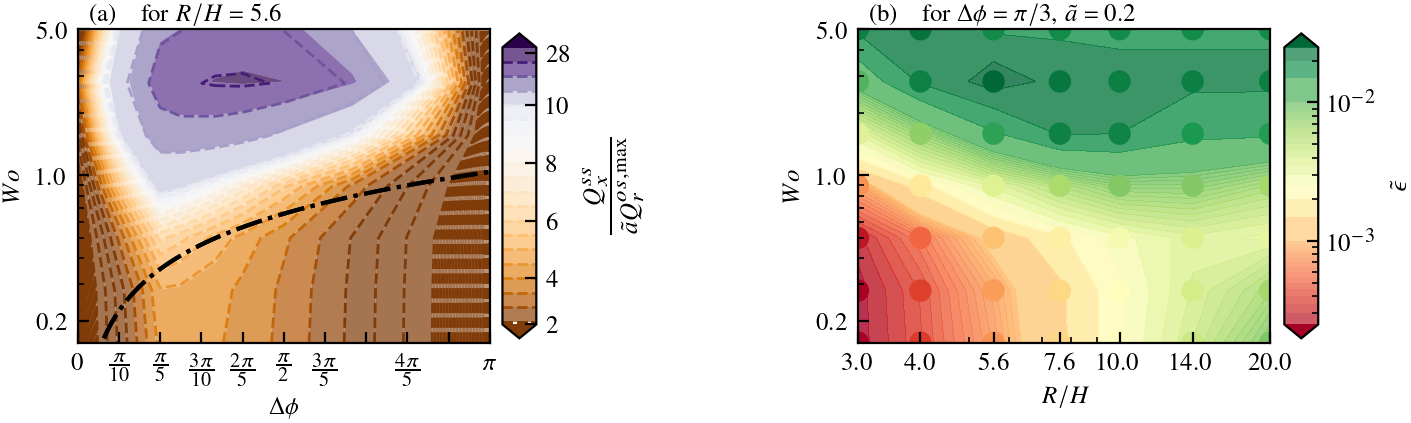}
	\caption{ \textbf{(a)} Ratio of steady axial to maximum oscillatory radial fluxes, rescaled by $\tilde{a}$. The overlaid contours for $\tilde{a}=0.1$ (dashed lines) and  $\tilde{a}=0.2$ (filled colour) demonstrate an excellent scaling collapse. \textbf{(b)} Contours of hydrodynamic efficiency ($\tilde{\epsilon}$) against channel half-depth ($R/H$) and Womersley number ($\Wo$) for $\tilde{a}=0.2$.}
	\label{fig:microfluidicOptim}
\end{figure}

\section{Geometric flux optimization}
\label{sec:optimization}

While pendular-wave motility is highly inefficient for bulk transport in biological systems, its underlying mechanism holds significant potential for engineered biomimetic applications. 
We explore a simple geometric strategy to maximize the pumped axial flux for a given dissipative energy loss.
Given the low hydrodynamic efficiency we deduce that oscillatory intervillous pumping plays a dominant role in determining dissipated power.
To characterize the overall strength of this oscillatory pumping, we take the peak value ($Q^{os,\max}_r$) of the time-averaged oscillatory radial flux per intervillous gap, $Q^{os}_r (r) = \frac{1}{2 T N} \int_{0}^{L_x} \int_{t}^{t+T} |u_r (t,x,r) - u^{ss}_r (x,r)| \, dt \, dx$, where steady field $u^{ss}_r (x,r) = \frac{1}{T} \int_{t}^{t+T} u_r (t,x,r) dt$.
In the Stokes flow regime, \citet{vernekar_HydrodynamicsVillipatternedChannel_2025} provide the geometric scaling that predicts the flux ratio $Q^{ss}_x/Q^{os,\max}_r \sim \tilde{a} W R \cos\left( \DPhi/2 \right) /(H P)$.
Figure~\ref{fig:microfluidicOptim}(a) confirms that rescaling the steady flux by $\tilde{a} Q^{os,\max}_r$ gives an excellent contour-collapse for both amplitudes.
Crucially, the contour landscape of rescaled flux ratio exhibits a strong qualitative correlation with the efficiency maps seen in figure~\ref{fig:pumpingEfficiencyRatio}, suggesting that $\left(\tilde{a} Q^{ss}_x /Q^{os,\max}_r\right)^2$ acts as a ``governing variable'' determining the axial pumping efficiency to leading order (also see supplementary figure~\SMfig{4}).

Thus, a viable geometric approach to augment the efficiency \textit{via} the flux ratio $Q^{ss}_x/Q^{os,\max}_r$ is to increase the half-channel depth $R$. 
We expect increasing $R$ to linearly increase the axial flux while leaving the radial flux largely unaffected.
Figure~\ref{fig:microfluidicOptim}(b) plots contours of the hydrodynamic efficiency $\tilde{\epsilon}$ over $3 \le R/H \le 20$ and $\Wo$, for fixed $\tilde{a}=0.2$ and the phase lag $\DPhi=\pi/3$.
In the viscous-dominated regime ($\Wo \lesssim 0.5$), efficiency monotonically increases, scaling quadratically with increasing $R/H$ thus validating our geometric optimization hypothesis (see supplementary figure~\SMfig{5}).
In this regime, the efficiency reaches a peak of $\tilde{\epsilon}=0.87\%$ at $R/H=20.0$ and $\Wo=0.16$.
However, as oscillatory fluid inertia increases ($\Wo > 1.0$), this geometric scaling breaks down.
The efficiency peak of $\tilde{\epsilon}=2.77\%$ occurs at $R/H = 5.6$ and $\Wo\approx 2.8$, before gradually declining to a plateau of $\approx2\%$ with further increase in $R/H$.
This scaling breakdown occurs because the onset of oscillatory inertia introduces a dynamically decreasing length scale (MBL, $\ell \propto \StLayer^{1/2}$) confining most of the driven flux into a narrow region above the villi tips.
This well-known Stokes layer effect ultimately renders any further increase in the channel height dynamically redundant.

\begin{table}
	\centering
	\begin{tabular}{lll c}
		\hline
		& Reference & \quad & Peak efficiency, $\tilde{\epsilon}$ \\ \hline
		&  \citet{osterman_FindingCiliaryBeating_2011} & 3D cilia array  &  ${\approx} 1.6\%$  \\
		&  \citet{elgeti_EmergenceMetachronalWaves_2013} & 3D cilia array  &  ${\approx} 1.4\%$  \\
		&  \citet{guo_EvaluatingEfficiencyRobustness_2016} & 3D individual cilium & ${\approx} 0.56\%$ \\
		&  Present work, inertial regime &  (for $R/H=5.6$) &  ${\approx} 2.77\%$ \\
		&  Present work, viscous regime & (for $R/H=20.0$) &  ${\approx} 0.87\%$ \\
	\end{tabular}
	\caption{Comparison of peak hydrodynamic efficiencies ($\tilde{\epsilon}$) with ciliary fluid propulsion.}
	\label{tab:eff_comparison}
\end{table}
\section{Discussion and conclusions}
\label{sec:conclusions}

Using simulations informed by isolated rat duodenum experiments, we investigate the hydrodynamic pumping efficiency in an idealized 2D channel geometry undergoing pendular-wave motility.
In the physiologically relevant viscous-dominated regime, the steady axial pumping efficiency is several orders of magnitude lower than that of peristalsis for an equivalent flux.
We therefore infer that bulk duodenal fluid pumping is not a physiologically significant function of propagating pendular motility.
Instead, we postulate that the primary physiological function of this propagating motility is the introduction of local shear into the secreted mucus barrier over the villi, thereby enhancing homogenisation and shear-induced nutrient diffusion.
Comparing the strain rate in the barrier region and the enstrophy density in the luminal region with those of canonical peristaltic reference values at equivalent wave speeds strongly supports this hypothesis.

With an eye on valveless flow-control applications, we demonstrated a simple optimization strategy of increasing the pumped flux by increasing the half-channel-to-villi ratio ($R/H$).
In the viscous-dominated regime, this strategy yields a quadratic efficiency increase ($\propto (R/H)^2$), reaching a peak value of $\tilde{\epsilon} \approx 0.87\%$ at $R/H=20$.
This low pumping efficiency for the villi-wall system ($\tilde{\epsilon} \lesssim 1\%$) is comparable, albeit markedly lower, to that for ciliary carpets undergoing optimal 3D metachronal motility in Stokes flow conditions (see table~\ref{tab:eff_comparison}).
Crucially, when driven into the inertial flow regime, our system reaches a peak efficiency of $2.77\%$, at $\Wo=2.8$, $\DPhi=\pi/3$ and $R/H=5.6$, nearly double that of the ciliary arrays.
However, in this inertial regime, the efficiency plateaus upon further increase in $R/H$ because of the Stokes layer effect that confines the bulk axial flux to a narrow zone above the villi tips, rendering any further channel expansion redundant.
Before drawing strict inferences from quantitative comparisons, we must note a significant distinction: the cited ciliary studies assume unbounded 3D half-domains, whereas we model a confined 2D channel.
If extended to a 3D rectangular channel with villous ridges, we expect the pumping efficiency to decrease due to additional viscous dissipation caused by 3D vortical flows \citep{hall_MechanicsCiliumBeating_2020}.

Notwithstanding the slightly lower efficiency compared to Stokes-regime ciliary pumping, the villi-patterned design offers advantages for biomimetic micro-pumping applications \citep{ni_RecentProgressDevelopment_2023, sun_OptimisationNetFlow_2025}.
Unlike artificial ciliary arrays that require complex, individually coordinated actuation patterns to achieve metachrony \citep{milana_MetachronalPatternsArtificial_2020, wang_ElectronicallyActuatedArtificial_2024}, these oscillating villi can be entirely rigid and driven by a simple travelling wave along the basal substrate.
This structural rigidity offers inherent benefits in fabrication simplicity and device durability compared to micro-pumps that rely on complex deformable actuators.
Furthermore, engineered systems can be deliberately driven at higher frequencies to induce inertial microfluidic effects, thereby accessing the higher pumping efficiencies.

\begin{acknowledgments}
	Most of the computations presented in this paper were performed using the GRICAD infrastructure (https://gricad.univ-grenoble-alpes.fr). LRP is part of the LabEx Tec21 (ANR-11-LABX-0030) and of the PolyNat Carnot Institute (ANR-11-CARN-007-01). The authors thank Agence Nationale de la Recherche for its financial support of the projects TransportGut, ANR-21-CE45-0015.
\end{acknowledgments}

\bibliography{../../UGA-TransportGut.bib}

\end{document}